\documentclass[aps,pre,reprint,floatfix,showkeys,longbibliography]{revtex4-1}	
\usepackage[utf8]{inputenc}     
\usepackage[english]{babel}   

\usepackage{lipsum}
\usepackage[hidelinks,colorlinks=true]{hyperref}

\usepackage[usenames,dvipsnames,svgnames,table]{xcolor}
\usepackage{graphicx}

\definecolor{myred}{RGB}{228,26,28}
\definecolor{myblue}{RGB}{55,126,184}
\definecolor{myorange}{RGB}{225,127,0}
\definecolor{mygreen}{RGB}{77,175,74}
\definecolor{mylila}{RGB}{152,78,163}
\definecolor{mybrown}{RGB}{153,76,0}
\definecolor{mygray}{RGB}{153,153,153}
\definecolor{darkred}{rgb}{0.8,0,0}
\definecolor{mydarkgreen}{RGB}{0,102,0}
\definecolor{mydarkbrown}{RGB}{102,52,0}
\definecolor{Orange}{RGB}{235,129,27}
\definecolor{Green}{RGB}{35,55,59}

\usepackage{bm}
\usepackage{amsmath,amsfonts,amssymb}

\usepackage{tikz} 
\usetikzlibrary{decorations,backgrounds,patterns} 
\usepackage{pgfplots} 
\usepackage{pgfplotstable}

\usepgfplotslibrary{external}
\pgfplotsset{compat=newest}
\usetikzlibrary{pgfplots.groupplots}
\usetikzlibrary{calc}
\usetikzlibrary{3d}
\usetikzlibrary{positioning}
\usepgfplotslibrary{units}
\pgfplotsset{unit code/.code={\si{#1}}}
\usepgfplotslibrary{patchplots}
\usepgfplotslibrary{fillbetween}
\tikzset{>=stealth}

\newcommand{\idest}{\emph{i.e. }}

\renewcommand{\d}{\mathrm{d}}
\newcommand{\D}{\partial}

\newcommand{\euler}{\mathrm{e}}
\renewcommand{\i}{\mathrm{i}}

\newcommand{\A}{\bm{A}}

\newcommand{\Real}{\operatorname{Re}}

\newcommand{\rates}{W}

\newcommand{\invtemperature}{\beta}

\newcommand{\force}{f}

\newcommand{\inforce}{f_1}

\newcommand{\outforce}{f_2}

\newcommand{\microsign}{\Theta(\microstate,\microstate')}
\newcommand{\mesosign}{\Theta(\mesostate,\mesostate')}
\newcommand{\meanfieldsign}{\Theta(i,j)}
\newcommand{\change}{\Delta}
\newcommand{\boltzmann}{k_b}
\newcommand{\Potential}{U}
\newcommand{\potential}{u}
\newcommand{\dimension}{N}
\newcommand{\occupation}{N}
\newcommand{\occupationdensity}{n}
\newcommand{\stateenergy}{\epsilon}
\newcommand{\microstate}{\alpha}
\newcommand{\mesostate}{\bm{N}}

\newcommand{\microprobability}{p}

\newcommand{\mesoprobability}{P}
\newcommand{\mesosteadyprobability}{P^{s}}

\newcommand{\meanfieldprobability}{\mathcal{P}}
\newcommand{\meanfieldcurrent}{\mathcal{I}}

\newcommand{\multiplicity}{\Omega}
\newcommand{\microrates}{w}
\newcommand{\mesorates}{W}

\newcommand{\meanfieldrates}{\mathcal{V}}

\newcommand{\arrheniusprefactor}{\Gamma}
\newcommand{\microenergy}{e}
\newcommand{\microheat}{q}
\newcommand{\microwork}{w}
\newcommand{\microentropy}{s}

\newcommand{\microep}{\sigma}

\newcommand{\mesoenergy}{E}
\newcommand{\mesoheat}{Q}
\newcommand{\mesowork}{W}
\newcommand{\mesoentropy}{S}
\newcommand{\mesoentropyflow}{S_e}
\newcommand{\mesoep}{\Sigma}
\newcommand{\internalentropy}{S^{int}}

\newcommand{\mesofreeenergy}{A}

\newcommand{\meanfieldenergy}{\mathcal{E}}
\newcommand{\meanfieldheat}{\mathcal{Q}}
\newcommand{\meanfieldwork}{\mathcal{W}}
\newcommand{\meanfieldpower}{\mathbb{P}}

\newcommand{\meanfieldentropy}{\mathcal{S}}

\newcommand{\singleworkcurrent}{ \langle\dot{\mesowork}_1^s \rangle }
\newcommand{\meanfieldworkcurrent}{ \overline{\meanfieldwork}}

\newcommand{\efficiency}{\eta}

\newcommand{\statenumber}{q}

\begin{document}
\title{Universality in driven Potts models}
\author{Tim Herpich}
\email{Electronic Mail: tim.herpich@uni.lu}
\author{Massimiliano Esposito}
\affiliation{Complex Systems and Statistical Mechanics, Physics and Materials Science Research Unit, University of Luxembourg, L-1511 Luxembourg, Luxembourg}
\date{\today}

\begin{abstract}
We study the stochastic dynamics of infinitely many globally interacting units made of $\statenumber$ states distributed uniformly along a ring that is externally driven. 
While repulsive interactions always lead to uniform occupations, attractive interactions give rise to much richer phenomena:
We analytically characterize a Hopf bifurcation which separates a high-temperature regime of uniform occupations from a low-temperature one where all units coalesce into a single state.
For odd $\statenumber$, below the critical temperature starts a synchronization regime which ends via a second phase transition at lower temperatures, while for even $\statenumber$ this intermediate phase disappears.  
We find that interactions have no effects except below critical temperature for attractive interactions. 
A thermodynamic analysis reveals that the dissipated work is reduced in this regime, whose temperature range is shown to decrease as $\statenumber$ increases. 
The $\statenumber$-dependence of the power-efficiency trade-off is also analyzed.
\end{abstract}
\maketitle

\section{Introduction}
\label{sec:intro}

While the thermodynamics of equilibrium phase transitions in interacting systems has a long history and is well-documented \cite{landau1994pergamon,yeomans1992oxford}, it is only as of recent that the thermodynamics of nonequilibrium phase transitions started to be explored \cite{herpich2018prx,verley2017epl,tomte2012prl,zhang2016jsm,crosato2018pre,crosato2018pre,gingrich2014pre,garrahanprl2014}. This delay can be attributed to the lack of a theory that systematically describes the thermodynamics of out-of-equilibrium processes. Over the past two decades, it has become evident that such a theory is embodied by stochastic thermodynamics that characterizes thermodynamic properties in systems exhibiting Markovian stochastic dynamics \cite{seifert2012rpp,broeck2015physica,zhang2012pr,sekimoto2010}. Here, the time-scales of the dynamics are specified by transition rates that incorporate the energetics of the nonequilibrium system via the so-called local detailed balance condition \cite{esposito2010njp,esposito2009rmp}. Stochastic thermodynamics has been successfully applied to characterize energy transduction in noninteracting or interacting few-body-systems \cite{seifert2012rpp,benenti2017pr,proesmans2016prx}. Recently, also interacting many-body thermodynamics and the relation between interactions and power-efficiency trade-offs have been investigated \cite{herpich2018prx, verley2017epl, imparato2015njp, imparato2012prl}.
Yet, the simplicity of these models allows to explicitly solve the dynamics. In this letter we demonstrate how stochastic thermodynamics allows to qualitatively capture a rich dynamical phenomenology of systems that are too complex to be addressed explicitly.

A popular model in statistical mechanics that exhibits an equilibrium phase transition is the Potts model \cite{ashkin1943pr} which generalizes the special case of the Ising model \cite{ising1925zfp,onsager1944pr} ($\statenumber=2$) by considering interacting spins on a lattice that can take $\statenumber$ different values distributed uniformly about a circle. While in equilibrium statistical mechanics this model has been largely explored \cite{wu1982rmp}, little is known about its out-of-equilibrium properties.
Progress in that direction was made in Ref. \cite{herpich2018prx} that studies an all-to-all interacting and driven three-state model (Potts model with $\statenumber=3$) across different scales.
This work generalizes the previous one by considering all-to-all interacting and driven Potts models at the mean-field level with variable $\statenumber$. The dynamics is modeled by thermodynamically consistent Arrhenius rates. We find that for repulsive interactions there is no phase transition. For attractive interactions we characterize via thermodynamic principles the distinct stationary states in the low- and high-temperature regime that are universal for all finite $\statenumber$. The key result is that we derive the $\statenumber$-dependent critical Hopf bifurcation temperature. Numerically, we show that for even $\statenumber$ this phase transition separates the high-temperature and the low-temperature phase, whereas for odd $\statenumber$ there are stable limit cycles implying the existence of a second phase transition. Finally, the dissipated work and the power-efficiency trade-off in the different phases are discussed.

\section{Model}
\label{sec:model}

We consider infinitely many units made of $\statenumber$ states with energies $\stateenergy_i$ ($i=1,2,\ldots \statenumber$) distributed uniformly along a ring in contact with a heat bath at inverse temperature $\invtemperature = (\boltzmann T)^{-1}$, where we set $\boltzmann \equiv 1$ in the following. Any unit in a given state $i$ interacts with all other units that occupy the same state $i$ with the global coupling constant $\potential $. The system is autonomously driven by a global and non-conservative force $\force$ that creates a bias along the ring with the rotational orientation $ 1 \to 2 \to \ldots \to \statenumber \to 1$, as depicted in the figure in appendix \ref{sec:appendix} that depicts a finite number of $\statenumber$-state units with $\statenumber=8$.
The system is fully characterized by the occupation densities $\meanfieldprobability_i$ of the states $i=1,\ldots,\statenumber$ that we identify as the probabilities for any unit in the mean-field to occupy these states.

We assume that the dynamics of the jump process is governed by a Markovian master equation (ME)
\begin{align}
\dot{ \meanfieldprobability }_i \! &= \sum\limits_{j } \meanfieldrates_{ij} \, \meanfieldprobability_j \, , \label{eq:meanfieldmasterequation}
\end{align}
where we choose Arrhenius transition rates from state $j$ to $i$, $ \meanfieldrates_{ij} = \arrheniusprefactor \, \exp[ -\frac{\invtemperature}{2}\! \left( \change \meanfieldenergy (i,j) - \meanfieldsign \force \right) ]$, with the kinetic prefactor $\Gamma$ and the function $\meanfieldsign$ that selects transitions between adjacent states along the ring according to their alignment with or against the bias $\force$ and that is defined as $\meanfieldsign=1$ for $ (i-j) \text{ mod } \statenumber = \! 1$ and $\meanfieldsign \! = \! - 1$ otherwise. Moreover, the change in energy due to that transition is given by $ \change \meanfieldenergy (i,j) = \stateenergy_i - \stateenergy_j + \potential(\meanfieldprobability_i - \meanfieldprobability_j) $. Thus, the transition rates satisfy the local detailed balance condition, \idest
\begin{align} \label{eq:meanfieldlocaldetailedbalance}
\ln \frac{ \meanfieldrates_{ij}}{ \meanfieldrates_{ji}} &= -\invtemperature \left( \stateenergy_i-\stateenergy_j + \potential(\meanfieldprobability_i-\meanfieldprobability_j)- \meanfieldsign \, \force \right) ,
\end{align}
ensuring that the system is thermodynamically consistent. We furthermore note that probability conservation, $\sum_i \meanfieldprobability_i =1$, removes one degree of freedom such that the system has $\statenumber-1$ dimensions. In the appendix \ref{sec:appendix}, the mean-field Eq. \eqref{eq:meanfieldmasterequation} is derived from a microscopic stochastic description for the many-body problem as an asymptotically exact Eq. in the limit of infinitely many units.

\section{Dynamics}
\label{sec:dynamics}

For a flat energy landscape of the units, $\stateenergy_i = \stateenergy \; \forall i$, the nonlinear Eq. \eqref{eq:meanfieldmasterequation} is solved by a uniform probability distribution
\begin{align} \label{eq:hightemperaturelimit}
\meanfieldprobability_i^s = 1/\statenumber , \quad  i =1,2,\ldots \statenumber .
\end{align}
Yet, no statements can be made about the stability of that fixed point without a stability analysis that for $\statenumber > 4$ is difficult.

We demonstrate that the thermodynamic consistency of the Potts model encoded in Eq. \eqref{eq:meanfieldlocaldetailedbalance} constrains the dynamics and even allows to generically specify the long-time solution in the low- and high-temperature regime. First, we note that the high-temperature limit $\invtemperature \to 0$ represents a reversible limit for finite $\force$ since detailed balance $\meanfieldrates_{ij} \, \meanfieldprobability_j^{eq} = \meanfieldrates_{ji} \, \meanfieldprobability_i^{eq}, \, \forall i,j$ holds. Equilibrium statistical mechanics prescribes that the system behaves entropically and thus the uniform probability distribution from Eq. \eqref{eq:hightemperaturelimit} represents a stable and unique fixed point. Next, the low-temperature limit, $\invtemperature \to \infty$ represents a totally irreversible limit for finite $\force$, where Eq. \eqref{eq:meanfieldmasterequation} reduces to $ \dot{ \meanfieldprobability }_i \! = \meanfieldrates_{i,i-1} \, \meanfieldprobability_{i-1} - \meanfieldrates_{i+1,i} \, \meanfieldprobability_{i}$. Here, the driving $\force$ is acting like a renormalization of the kinetic prefactor. This suggests that occupation is the only relevant factor that determines where the dynamics goes to in the long-time limit, \idest the system with irreversible rates behaves energetically like an equilibrium one would.
In this case one has to distinguish between repulsive ($\potential > 0$) and attractive ($\potential <0$) interactions. For the former the system has a unique energy ground state that coincides with the entropic state from Eq. \eqref{eq:hightemperaturelimit}. Conversely, for attractive interactions there are $\statenumber$ energy ground states where all units occupy the same state
\begin{align} \label{eq:lowtemperaturelimit}
\meanfieldprobability_i^s =1,\; \meanfieldprobability_{j\neq i}^s=0, \quad i =1,2,\ldots,\statenumber .
\end{align}
These states indeed correspond to fixed points as can be readily verified by inserting Eq. \eqref{eq:lowtemperaturelimit} into the ME \eqref{eq:meanfieldmasterequation}. To which of those fixed points the dynamics is striving depends on which state $i$ is initially the most populated one. We emphasize that these results hold for any finite number of states $\statenumber$ and finite autonomous driving $\force$.

At all temperatures, the stability of the symmetric fixed point, $\meanfieldprobability_i^* \equiv 1/\statenumber$, is encoded in the spectrum of the linearized Jacobian, $ \A_{ij} \!\!\! ~ \equiv ~ \!\!\! \sum_{k} [ \D ( \meanfieldrates_{ik} \meanfieldprobability_k) / \D \meanfieldprobability_j ] |_{ \bm{\meanfieldprobability}^*} $. For $\statenumber \leq 4$ we find that the critical point $\invtemperature_{c}$ obeys the relation
\begin{align} \label{eq:criticalpoints}
( \statenumber + \invtemperature_{c} \potential ) = 0  .
\end{align}
For repulsive interactions this relation is never satisfied and the system remains in the symmetric fixed point at all temperatures.

A key finding is that for attractive interactions the dynamics has a rich phenomenology: The main result is that we prove that relation \eqref{eq:criticalpoints} also characterizes the critical temperature for $\statenumber > 4$. To this end, we evaluate the linear stability matrix for $\statenumber > 2$ at the critical temperature \eqref{eq:criticalpoints} and obtain
\begin{align} \label{eq:stabilitymatrixcriticalpoint}
\A_{ij} &= \! c \big[ (\delta_{i+1,j} \big|_{i\neq \statenumber} - \delta_{i-1,j} \big|_{i\neq 0} ) \! + \delta_{i,q} \, \delta_{0,j} - \delta_{i,0} \, \delta_{q,j} \big]  ,
\end{align}
with $ c = \Gamma \sinh \left( \statenumber \force/(2 \potential) \right) $. This skew-symmetric circulant matrix admits the eigenvalues \cite{davis1970wiley}
\begin{align}
\lambda_k &= 2 \i \, c \sin \left( k \frac{2 \pi}{\statenumber} \right) ,  \qquad 
k=0,1,\ldots,\statenumber - 1 ,
\end{align}
that are thus either identical zero or purely imaginary.
For $\statenumber =2$ the linear stability matrix, $A_{ij}=0$, has only zero eigenvalues. 
Next, we consider temperatures in the vicinity of the critical temperature, $\invtemperature_{c} + \delta \invtemperature$, and expand the linear stability matrix in Eq. \eqref{eq:stabilitymatrixcriticalpoint} up to linear order, $\A(\invtemperature_{c} + \delta \invtemperature) = \A(\invtemperature) \big|_{\invtemperature_c} + \delta \invtemperature \; \bm{B} + \mathcal{O}\big(\delta \invtemperature^2 \big) $,
with the first-order correction matrix
\begin{align}
\bm{B}_{ij} &= a \delta_{ij} + b( \delta_{i+1,j} + \delta_{i,q} \delta_{0,j} ) + c ( \delta_{i-1,j} + \delta_{i,0} \, \delta_{q,j} ) \nonumber \\
a &= \!-\! \tfrac{2 \Gamma \potential }{\statenumber} d, \,
b = \Gamma \tfrac{ 2 \potential \!-\! \statenumber \force}{2 \statenumber} d , \,  
c =  \Gamma \tfrac{\statenumber \force \!+\! 2 \potential }{2 \statenumber} d, \,
d \!=\! \cosh \big( \statenumber  \tfrac{ \force}{2 \potential} \big) \nonumber , 
\end{align}
which has the eigenvalues
\begin{align}
\tilde{\lambda}_k \!= \!  - \tfrac{2 \Gamma d  \; \sin^2 \! \big( \tfrac{k \pi}{\statenumber} \big) } {\statenumber } \! \bigg[ 2 \potential \!+\! \i \force \statenumber \cot \big( \tfrac{k \pi}{\statenumber} \big) \! \bigg] \!, \, k=0,1,\ldots,\statenumber \!-\! 1
.  \nonumber
\end{align}
We note that for all $\statenumber$, $\Real [\tilde{\lambda}_k ] \geq \!0 \; \forall k$, such that the real parts of all perturbed eigenvalues, $  \Real [ \lambda_k ] + \delta \invtemperature \, \Real[ \tilde{\lambda}_k ] $, change their sign at the critical point from negative ($\delta \invtemperature < 0$) to positive values ($\delta \invtemperature > 0$) or remain zero.

\begin{figure}[h!]
\begin{center}

\includegraphics[scale=1]{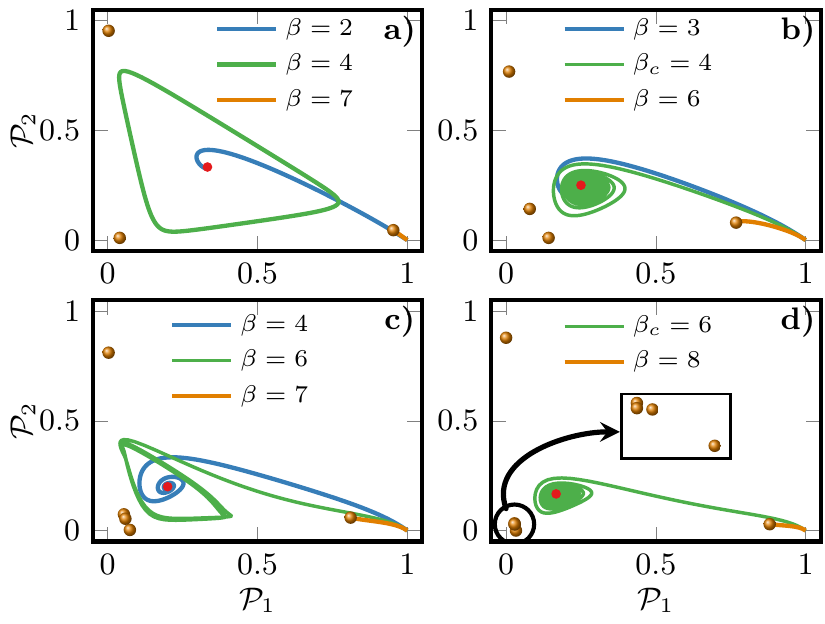}

\caption{Parametric plot of the probabilities $ \meanfieldprobability_{1,2} $ for $\statenumber=3$ [panel a)], $\statenumber=4$  [panel b)], $\statenumber=5$  [panel c)] and $\statenumber=6$  [panel d)] for different temperatures $\invtemperature$. The orange-shaded spheres indicate the $\statenumber$ asymmetric fixed points and the symmetric fixed point is represented by the red closed circle. In all plots the initial condition $\meanfieldprobability_1(0)=1$ is used and the chosen values for the dimensionless parameters read $\Gamma=0.1$ and $\force=1$. \label{fig:dynamics} }
\end{center}
\end{figure}

Thus Eq. \eqref{eq:criticalpoints} indeed characterizes the critical point that destabilizes the symmetric fixed point. If the system is at equilibrium, $\force = 0$, the real parts of the eigenvalues of the linear stability matrix evaluated at the critical point change their sign or remain zero while the imaginary parts are identically zero corresponding to a saddle-node bifurcation that degenerates the single symmetric fixed point into multiple fixed points. In the out-of-equilibrium scenario, $\force \neq 0$, the real parts of the eigenvalues of the linearized Jacobian evaluated at the critical temperature change their sign or remain zero while the imaginary parts remain finite implying a Hopf-bifurcation that degenerates the symmetric fixed point into a limit cycle.
Equation \eqref{eq:criticalpoints} states that the uniform probability distribution can be observed over a larger range of temperatures as $\statenumber$ increases. The uniform distribution, however, removes the interactions from the dynamics [cf. Eq. \eqref{eq:meanfieldlocaldetailedbalance}]. Hence for repulsive interactions the mean-field system is noninteracting as well as for attractive interactions at temperatures above the first critical temperature that approaches zero as $\statenumber$ becomes large. In the following we consider attractive interactions ($\potential =-1$) if not explicitly stated otherwise.
To infer the stability of the limit cycles, an analysis of the normal form of the Hopf bifurcation and the computation of the first Lyapunov coefficient would be required \cite{kuznetsov1998}, which for $\statenumber > 3$ renders analytic progress difficult. A numerical analysis in Fig. \ref{fig:dynamics} depicting in a parametric $\meanfieldprobability_1-\meanfieldprobability_2$ plot the dynamics of the $\statenumber$-model ($\statenumber=3,4,5,6$) motivates the following conjecture:

If $\statenumber$ is even like in panels b) and d), the Hopf bifurcation occurs subcritical, \idest the limit cycle is unstable and degenerates into $\statenumber$ asymmetric stable fixed points, hence there is only one phase transition at $\invtemperature_c$. Conversely, if $\statenumber$ is odd, like in panels a) and c), the Hopf-bifurcation occurs supercritical, \idest the symmetric fixed point degenerates at $\invtemperature_{c_1}$ into a stable limit cycle. Physically, a limit cycle in the $\statenumber$-dimensional probability space means that the units tend to undergo the same transition together at a given time, i.e. they synchronize. Since the low-temperature limit in Eq. \eqref{eq:lowtemperaturelimit} is also satisfied for odd $\statenumber$, there is a second critical point $\invtemperature_{c_2}$ at which the limit cycle degenerates via an infinite-period bifurcation \cite{keener1981siam} into $\statenumber$ asymmetric stable fixed points. In both cases, the multiple fixed points are related to each other by permutations of their coordinates in the $\statenumber$-dimensional probability space.
For decreasing temperatures these fixed points move towards the respective energy ground states in Eq. \eqref{eq:lowtemperaturelimit}. 
Thus, we have demonstrated that there are two classes of universal phenomenology: For all $\statenumber$ the high-temperature (low-temperature) regime is characterized by a single (multiple) (a)symmetric stable fixed point(s), while only for odd $\statenumber$ there is also an intermediate phase exhibiting stable oscillations. This universality is robust to slight changes of the state energies $\epsilon_i$. For large changes, the critical phenomena vanish and there is a single stable fixed point at all temperatures.
We furthermore emphasize that for sufficiently high-dimensional lattices and large system sizes, finite-range interactions will also generate the above phenomenology as discussed in Ref. \cite{herpich2018prx} and explicitly demonstrated in Ref. \cite{wood2006prl}. The choice of all-to-all interactions however allows to analytically characterize the universal phenomenology.

It is natural to consider the large-$\statenumber$ limit where the Potts model becomes the XY model. According to Eq. \eqref{eq:criticalpoints}, the phase in which the system exhibits a unique symmetric fixed point and thus is noninteracting exists over an increasing range of temperatures. Hence for exceedingly large $\statenumber$ the differences between systems with attractive interactions (for both even or odd $\statenumber$) and repulsive interactions at experimentally meaningful temperatures vanish.

\section{Thermodynamics}
\label{sec:thermodynamics}

Our thermodynamically consistent formulation of the Potts model [cf. Eq. \eqref{eq:meanfieldlocaldetailedbalance}] allows us now to address its nonequilibrium thermodynamic properties. In the appendix \ref{sec:appendixstochasticthermodynamics} a thermodynamic description is systematically established. The first law of thermodynamics 
\begin{align}
\d_t \, \meanfieldenergy  = \sum\limits_{i,j} \meanfieldenergy_{i} \, \meanfieldrates_{i j} \, \meanfieldprobability_{j}   
=  \dot{\mathcal{\mesoheat}} +  \dot{\mathcal{\mesowork}}  ,
\end{align}
states that the rate of change in internal energy $\meanfieldenergy$ is balanced by heat and work currents, $\dot{\meanfieldheat}$ and $\dot{\meanfieldwork}$, hence ensuring energy conservation. 

Furthermore, the non-negativity of the entropy production
\begin{align}
\dot{\meanfieldentropy}_i = - \d_t \sum_j \meanfieldprobability_j \ln \meanfieldprobability_j - \invtemperature \dot{\meanfieldheat} \geq 0 ,
\end{align}
constitutes the second law of thermodynamics. Hence, in the long-time limit, the entropy production, up to temperature, is equal to the work, $\dot{\meanfieldentropy}_i = \invtemperature \dot{\meanfieldwork}$.

We proceed by demonstrating that the bifurcations translate into nonequilibrium phase transitions that can be characterized via the work. Here, the work is dissipative since the mean-field system takes rotational energy, $ \dot{\meanfieldwork} >0 $, and dissipates it into the bath in form of heat, $ \dot{\meanfieldheat} <0 $. First, we recall that except for attractive interactions below the critical temperature in Eq. \eqref{eq:criticalpoints} the system behaves like a noninteracting one. The stationary average dissipated work current for a single unit reads
\begin{align}  \label{eq:singleunitworkhightemperature} 
\singleworkcurrent = 2 \Gamma  \force \sinh \left( \frac{\invtemperature \force}{2}\right) \geq 0, 
\end{align}
and is independent of the number of states $\statenumber$. Next, for $\invtemperature \gg \invtemperature_{c(c_2)} $, the stationary mean-field work current can be approximated as
\begin{align} 
\dot{\meanfieldwork}^s &\approx 2 \Gamma \, \force \, \euler^{\frac{\invtemperature \potential}{2}}  \sinh \left( \tfrac{ \invtemperature \force}{2} \right) = \euler^{\frac{\invtemperature \potential}{2}} \, \langle\dot{\mesowork}_1^s \rangle . \label{eq:meanfieldunitworklowtemperature}
\end{align}
Hence operating an interacting system in the low-temperatures regime is exponentially less costly in the interaction strength than maintaining a noninteracting one.

\begin{figure}[h!] 
\begin{center}

\includegraphics[scale=1]{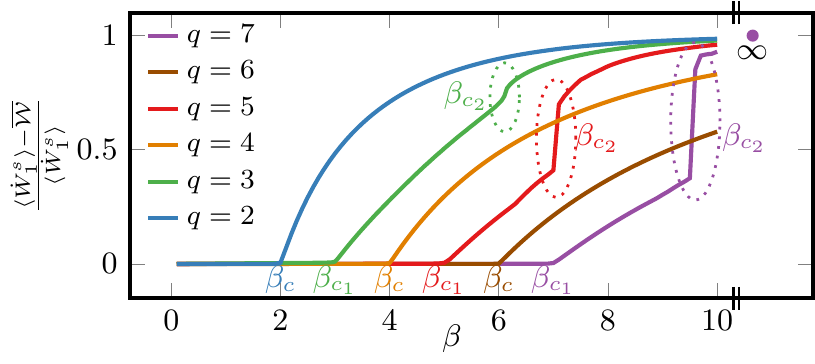}

\caption{Difference of the dissipated work for a stationary single-unit, $ \singleworkcurrent $, and the time-averaged mean-field work current $\meanfieldworkcurrent$ for inverse temperatures $\invtemperature = 0,1, \ldots ,10$ and different $\statenumber=2,3, \ldots,7$. The purple closed circle represents the analytic result in the low-temperature limit. \label{fig:dissipatedwork}  }
\end{center}
\end{figure}

This can be seen in Fig. \ref{fig:dissipatedwork} that depicts the difference between the stationary work current of a single unit $\singleworkcurrent$ and the asymptotic mean-field work current $\meanfieldworkcurrent$ as a function of $\invtemperature$ for different $\statenumber$.
In agreement with Eqs. \eqref{eq:singleunitworkhightemperature} and \eqref{eq:meanfieldunitworklowtemperature}, we find that for all $\statenumber$ the mean-field system is noninteracting at inverse temperatures below the inverse Hopf-bifurcation temperature $\invtemperature_{c(c_1)}$ [Eq. \eqref{eq:criticalpoints}], while the dissipated work is significantly reduced above that inverse critical temperature. In fact, we conclude from the monotonotic behavior of the curves that it is always energetically beneficial to consider attractive interactions. Since the (first) critical point is shifting to larger values of $\invtemperature$ as $\statenumber$ increases, it is overall favorable to employ small-$\statenumber$ units. At the (first) critical point $\invtemperature_{c(c_1)}$ the mean-field dissipated work exhibits for all $\statenumber$ a kink that is reminiscent of a first-order equilibrium phase transition. As a consequence of the two bifurcations for odd $\statenumber$ there is a second non-equilibrium phase transition at $\invtemperature_{c_2}(\force)$ which displays characteristics of both a saddle and a jump that is more pronounced for larger $\statenumber$.

A central result of this work is that small and odd-$\statenumber$ units not only dissipate less when interacting attractively, but also allow to realize the optimal power-efficiency trade-off in energy transduction. To construct an energy converter, we decompose the nonconservative force $\force$ into a force aligned with ($\inforce > 0$) and a force acting against ($\outforce < 0$) the bias, \idest $\force = \inforce + \outforce $. We denote by $\mathcal{\meanfieldcurrent}$ the current aligned with the net-force $\force$ and define the input $\dot{\meanfieldwork}_1 \equiv \inforce \meanfieldcurrent$ and the output work current $\dot{\meanfieldwork}_2 \equiv \outforce \meanfieldcurrent$.

This work-to-work conversion is a commonly used concept to model energy transduction in molecular motors such as kinesin and myosin \cite{julicher1997rmp,imparato2015njp}.
For practical purposes the efficiency at maximum power (EMP) \cite{curzon75ajp} is of particular interest. The EMP is obtained by first maximizing the asymptotic output power $\overline{ \meanfieldpower } \equiv \dot{\meanfieldwork}_2$ with respect to $\outforce$. Next, in the long-time limit the efficiency is defined as $\efficiency \equiv -\meanfieldwork_2/\meanfieldwork_1 = - \outforce/\inforce \leq 1 $ and evaluated at maximum power.

\begin{figure}[h!] 
\begin{center}

\includegraphics[scale=1]{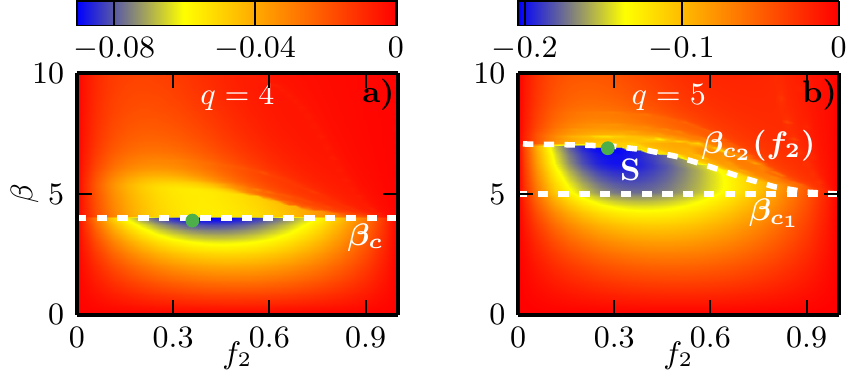}

\caption{The time-averaged output power $\overline{\meanfieldpower}$ as a function of $\outforce$ and $\invtemperature$ for $\statenumber = 4$ [panel a)] and $\statenumber = 5$ [panel b)]. The simulation time is chosen sufficiently large to ensure convergence of the time-averaged output power to its asymptotic value.
In both plots the white dashed lines indicate the set of critical points, hence the enclosed area defines the synchronization phase S. In addition, the global maximum of the output power is indicated by the green closed circle.
\label{fig:poweroutput} }
\end{center}
\end{figure}

Fig. \ref{fig:poweroutput} shows the asymptotic power output $\overline{\meanfieldpower}$ as a function of $\invtemperature$ and $\outforce$ for $\statenumber=4$ [panel a)] and $\statenumber=5$ [panel b)].
We first recall that in both cases for inverse temperatures below $\invtemperature_{c(c_1)}$ the systems are noninteracting and their power output is thus determined via Eq. \eqref{eq:singleunitworkhightemperature}. Next, the power rapidly drops in the low-temperature regime, that is for $\invtemperature > \invtemperature_c$ [panel a)] or $\invtemperature > \invtemperature_{c_2}(\outforce) $ [panel b)]. Thus, we find that the maximum power is achieved in the synchronization regime, that is for odd $\statenumber$-systems.

\begin{figure}[h!]
\begin{center}

%
%
%
%
%
%

\includegraphics[scale=1]{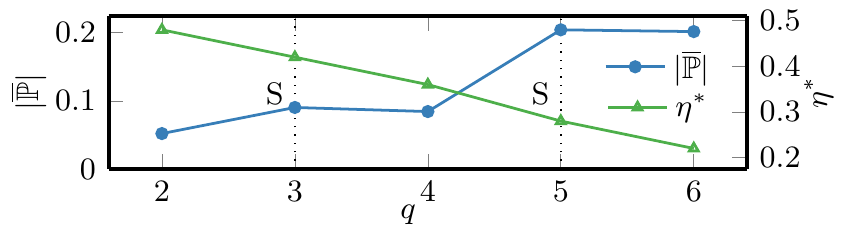}

\caption{The modulus of the time-averaged global maximum power $|\overline{\meanfieldpower}|$ and the associated EMP for different $\statenumber$. The black dotted lines indicate the $\statenumber$-values that exhibit a synchronization (S) phase.
\label{fig:powerefficiencytradeoff} }
\end{center}
\end{figure}

Fig. \ref{fig:powerefficiencytradeoff} depicts both the asymptotic global maximum power $| \overline{\meanfieldpower} |$ - indicated by green closed circles in Fig. \ref{fig:poweroutput} - and the EMP as a function of $\statenumber$.
\noindent Overall, the maximum power output is increasing with $\statenumber$, though there are striking jumps from even to odd $\statenumber$-systems, \idest from asynchronous to synchronized systems. These jumps are followed by plateaus where the maxima remain roughly the same. Conversely, the associated EMPs are decreasing monotonically with $\statenumber$. In particular, for $\statenumber \leq 3$, the EMP is close to the optimal value $1/2$ that is universal for a system with a single net-current responding linearly \cite{vandenbroeck2005prl,esposito2009prl}. Therefore, we conclude that the optimal power-efficiency trade-off is achieved for small and odd-$\statenumber$ systems that are compatible with synchronization.

\section{Conclusion}
\label{sec:conclusion}

Our work makes progress in the (thermodynamic) study of interacting systems away from equilibrium - a topic that is still in its infancy and for which general results are completely lacking.
The methods developed show that combining thermodynamic arguments with tools from nonlinear dynamics can help to characterize the complex dynamical behaviors of interacting systems described by stochastic thermodynamics without explicitly solving the dynamics. While our present study used Arrhenius rates, our methods are generic and applicable to any rates satisfying local detailed balance in Eq. \eqref{eq:meanfieldlocaldetailedbalance} (e.g. Glauber dynamics \cite{levin,yang1992prb}). 
Our findings also show that engineering interactions among collections of Brownian machines is a promising strategy to improve their performance.

\section{Acknowledgments}

T. H. thanks Artur Wachtel for fruitful interactions. The simulations were carried out using the HPC facilities of the University of Luxembourg \cite{hpc}. We acknowledge support by the National Research Fund, Luxembourg, in the frame of the AFR PhD Grant 2016, No. 11271777 and by the  European Research Council project NanoThermo (ERC-2015-CoG Agreement No. 681456).

\appendix

\section{Stochastic Dynamics on different Scales}
\label{sec:appendix}
\subsection{Many-Body Model}
\begin{figure}[h!]
  \centering
\includegraphics[scale=1]{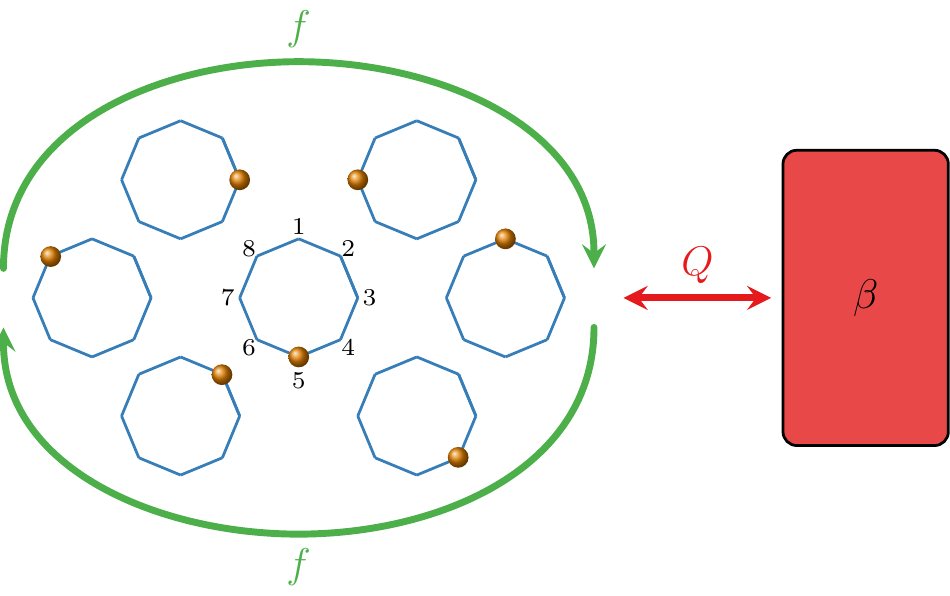}
\caption{Schematics of identical and globally interacting eight-state units that are connected with a heat bath at inverse temperature $\invtemperature$ and subjected to a nonconservative rotational force $\force$.  \label{fig:pottsmodelnetwork} }
\end{figure}

The following is devoted to establishing a thermodynamically consistent and microscopic many-body description of the Potts model for a finite number of units $\dimension$. From this representation of the Potts model we rigorously derive the mean-field equation \eqref{eq:meanfieldmasterequation} if the macroscopic limit $\dimension \to \infty$ is taken. Fig. \ref{fig:pottsmodelnetwork} illustrates the setup for a many-body eight-state model ($\statenumber$ =8). We want to remark that the following procedure is to a large extent analogous to the one detailed in Secs. II and VI of Ref. \cite{herpich2018prx} that studies in great detail the Potts model for $\statenumber =3$.

\subsection{Microscopic Dynamics}
We consider $\dimension$ $\statenumber$-state units with energies $\epsilon_i$, $i = 1,2,\ldots, \statenumber$. All units that occupy the same state $i$ interact globally with the coupling constant $\potential/\dimension$. The microscopic dynamics of the $\dimension$-body system is fully characterized by a microstate $\microstate$ which we denote by a multiindex $\microstate~=~(\microstate_1,\ldots,\microstate_i,\ldots,\microstate_{\dimension})$ with $\microstate_i = 1,2,\ldots \statenumber$. As an expample, we consider a transition from $\microstate '$ to $\microstate$ corresponding to a change in single-unit energy $\stateenergy_j \to \stateenergy_i$. Hence the occupation numbers change along that transition as $\occupation_j \to \occupation_j -1$ and $\occupation_i \to \occupation_i +1$. In order to determine the change in internal energy for this transition the interaction energy $\Potential(\microstate)$ of the network is required. One has
\begin{align}
 \Potential(\microstate) = \frac{u}{N} \sum\limits_{k=1}^{\statenumber} \; \sum\limits_{l=1}^{\occupation_{k}\!(\microstate)\!-\!1} \! l 
 = \frac{\potential}{2 \dimension} \sum\limits_{k=1}^{\statenumber} \occupation_k^2(\microstate) + \Potential_0 ,
\end{align}
thus the change in internal energy reads
\begin{align}
	\change \mesoenergy (\microstate,\microstate') = \stateenergy_{i}-\stateenergy_{j} + \frac{\potential}{\dimension} (\occupation_i - \occupation_j + 1) ,
\end{align}
where $\occupation_k^2(\microstate)$ denotes the number of units occupying the same single-unit state for the given microstate $\microstate$. 
We assume that the dynamics of the jump process is governed by a Markovian master equation (ME)
\begin{align} \label{eq:micromasterequation}
  \dot{\microprobability}_{\microstate}= \sum\limits_{ \microstate' } \microrates_{\microstate\microstate'} \, \microprobability_{\microstate'} \, ,
\end{align}
where $\microprobability_{\microstate}$ is the probability to be in the microstate $\microstate$ and the microscopic transition rates read
\begin{align} 
\microrates_{\microstate\microstate'} &= \arrheniusprefactor \; \euler^{-\frac{\invtemperature}{2} \left( \change E(\microstate,\microstate') - \microsign \force \right)}  \, . \label{eq:microtransitionrates}
\end{align}
The function $\microsign$ characterizes the transitions according to their alignment with the bias $f$. It is defined as $\microsign\!=\!1$ for $ \sum_i ( \microstate_i \! - \! \microstate'_i ) \! = \! 1$ mod $\statenumber$ and $\microsign \! = \! - 1$ otherwise. The microscopic transition rates satisfy local detailed balance
\begin{align} \label{eq:microlocaldetailedbalance}
\ln \frac{\microrates_{\microstate\microstate'}}{\microrates_{\microstate'\microstate}} = - \invtemperature \left( \change E(\microstate,\microstate') - \microsign \force \right)  \, ,
\end{align}
and thus ensure that the microscopic system is thermodynamically consistent.

\subsection{Mesoscopic Dynamics}
\label{sec:coarsegrainingdynamics}
In the microscopic formulation the system has a state space that grows exponentially with the number of units in the network as $\statenumber^{\dimension}$. However, the all-to-all interactions allow to determine the energy changes in the system without knowledge of the network topology, hence the microscopic ME (\ref{eq:micromasterequation}) for the full microstate dynamics can be coarse-grained as follows
\begin{subequations}
\begin{align} \label{eq:coarsegrainingscheme}
\dot{\mesoprobability}_{\mesostate} &= \sum\limits_{\microstate \in \mesostate} \sum\limits_{\mesostate'} \sum\limits_{\microstate' \in \mesostate'} \microrates_{\microstate\microstate'} \, \microprobability_{\microstate'} \\
&= \sum\limits_{\mesostate'} \microrates_{\mesostate \mesostate'}  \sum\limits_{\microstate \in \mesostate} \sum\limits_{ \microstate' \in \mesostate' } \; \microprobability_{\microstate'} \, \chi_{\microstate',\microstate}  \\
&= \sum\limits_{\mesostate'} \rates_{\mesostate \mesostate'}\; \mesoprobability_{\mesostate'} \, , \label{eq:mesomasterequation}
\end{align}
\end{subequations}
where $\mesoprobability_{\mesostate} \equiv  \sum_{\microstate \in \mesostate } \, \microprobability_{\microstate} $ refers to the mesoscopic probability to observe a mesostate $\mesostate \equiv (\occupation_1,\ldots,\occupation_{\statenumber})$ that comprises all microstates $\microstate \in \mesostate$. The mesoscopic transition rates are given by $\mesorates_{\mesostate \mesostate'} = \multiplicity(\mesostate,\mesostate') \microrates_{\mesostate \mesostate'}$ with the microscopic transition rates in mesoscopic representation $ \microrates_{\mesostate \mesostate'} $. Moreover, the characteristic function $\chi_{\microstate',\microstate} = 1$ if transitions between $\microstate'$ and $\microstate$ are possible and is 0 otherwise, hence one has
\begin{align}
\multiplicity(\mesostate,\mesostate') = \occupation'_1\, \delta_{\occupation'_1,\occupation_1+1} + \ldots +  \occupation'_{\statenumber} \, \delta_{\occupation'_{\statenumber},\occupation_{\statenumber}+1} \, .
\end{align}
We identify the Boltzmann entropy as the mesoscopic equilibrium entropy
\begin{align} \label{eq:internalentropy}
\internalentropy(\mesostate) = \ln \multiplicity(\mesostate) \, ,
\end{align}
where $ \multiplicity(\mesostate) = \dimension!/\prod_i \occupation_i! $ gives the number of microstates $\microstate$ that belong to the mesostate $\mesostate$. Collecting results, the mesoscopic local detailed balance relation is given by
\begin{align} \label{eq:mesolocaldetailedbalance}
\ln \frac{ \rates_{\mesostate\mesostate'} }{  \rates_{\mesostate'\mesostate} } = - \invtemperature \left[\change \mesofreeenergy(\mesostate,\mesostate')  - \mesosign \force \right] ,
\end{align}
where $\change \mesofreeenergy(\mesostate,\mesostate') \!=\! \Delta \mesoenergy(\mesostate,\mesostate')  - \! \invtemperature^{-1} \change \internalentropy(\mesostate,\mesostate') $ denotes the change in equilibrium free energy between the mesostates $\mesostate'$ and $\mesostate$. The function $\mesosign$ is defined as in the microscopic case in Eq. \eqref{eq:microlocaldetailedbalance}. Hence, $\mesosign=1$ if $ (i-j) \text{ mod } \statenumber = \! 1$ for the transition $ \mesostate_i' \to \mesostate_i, \mesostate_j' \to \mesostate_j$ and $\mesosign \! = \! - 1$ otherwise. We emphasize that the exact coarse-graining of the microscopic dynamics significantly reduces the complexity of the problem since the state space in the mesoscopic representation is growing like $\dimension^{\statenumber-1}/(\statenumber-1)! \,$.
\subsection{Mean-Field Limit}
\label{sec:coarsegrainingdynamics}
The asymptotic solution in the macroscopic limit ($\dimension \to \infty$) is simply given by a mean-field approximation performed on the mesoscopic ME \eqref{eq:mesomasterequation} scaled with $1/\dimension$ . The mean-field approximation amounts to replacing any $n$-point correlation function with a product of $n$ averages and thus yields
\begin{align}  \label{eq:meanfieldmasterequationderivation}
\meanfieldprobability_i \equiv \langle \dot{\occupationdensity}_i \rangle &= \sum\limits_{j} \mesorates_{ij} \big(\langle \occupationdensity_i \rangle ,\langle \occupationdensity_j \rangle \big) \equiv 
\sum\limits_{j} \meanfieldrates_{ij} \big( \meanfieldprobability_i , \meanfieldprobability_j  \big)  \meanfieldprobability_j  , 
\end{align}
which indeed is the nonlinear mean-field equation Eq. \eqref{eq:meanfieldmasterequation} . Here, we introduced the mean-field rates $\meanfieldrates_{ij} = \Gamma \exp(\epsilon_i-\epsilon_j + \potential( \meanfieldprobability_i - \meanfieldprobability_j ) - \meanfieldsign \force ) $ with the sign function defined as $\meanfieldsign=1$ for $ (i-j) \text{ mod } \statenumber = \! 1$ and $\meanfieldsign \! = \! - 1$ otherwise.

\section{Thermodynamic Laws on different Scales}
\label{sec:appendixstochasticthermodynamics}
\subsection{Microscopic Thermodynamics}
We start with the elementary thermodynamic state functions in this model: the microscopic internal energy and the microscopic system entropy
\begin{align} \label{eq:microinternalenergy}
\left\langle \microenergy \right \rangle &= \sum\limits_{\microstate} \mesoenergy(\microstate) \, \microprobability_{\microstate} \\
\left\langle \microentropy \right \rangle &= - \sum\limits_{\microstate} \microprobability_{\microstate} \ln \microprobability_{\microstate} \, . \label{eq:microsystementropy}
\end{align}
The time-derivative of the internal energy
\begin{align}
\d_t \langle \microenergy \rangle = \sum\limits_{\microstate, \microstate'} \mesoenergy(\microstate) \microrates_{\microstate \microstate'}  \, \microprobability_{\microstate'} 
= \langle \dot{\microheat} \rangle + \langle \dot{\microwork} \rangle , \label{eq:microfirstlaw}
\end{align}
stipulates the microscopic first law of thermodynamics that ensures energy conservation. Here, we defined the heat and work current as follows
\begin{align}
\langle \dot{\microheat} \rangle &= \sum\limits_{\microstate, \microstate'} \left[ \mesoenergy(\microstate) - f \,\microsign  \right] \microrates_{\microstate \microstate'}  \, \microprobability_{\microstate'} \\
\langle \dot{\microwork} \rangle &= \sum\limits_{\microstate, \microstate'} f \; \microsign \; \microrates_{\microstate \microstate'}  \, \microprobability_{\microstate'} \, ,
\end{align}
where the sign function $\microsign$ is equal to the one introduced below Eq. (\ref{eq:microlocaldetailedbalance}). The rate of change of the system entropy 
\begin{align} \label{eq:microentropybalance}
\d_t \langle \microentropy \rangle = \langle \dot{\microentropy}_e \rangle + \langle \dot{\microep} \rangle
\end{align}
can be decomposed into the entropy flow from the bath to the system
\begin{align} \label{eq:microentropyflow}
\langle \dot{\microentropy}_e \rangle = -  \sum\limits_{\microstate, \microstate'} \microrates_{\microstate \microstate'} \, \microprobability_{\microstate'} \ln \frac{\microrates_{\microstate \microstate'}}{\microrates_{\microstate' \microstate}}
= \invtemperature \langle \dot{\microheat} \rangle ,
\end{align}
and the non-negative entropy production rate
\begin{align} \label{eq:microsecondlaw}
\langle \dot{\microep} \rangle =  \sum\limits_{\microstate, \microstate'} \microrates_{\microstate \microstate'} \, \microprobability_{\microstate'} \, \ln \frac{\microrates_{\microstate \microstate'} \microprobability_{\microstate'} }{\microrates_{\microstate' \microstate} \microprobability_{\microstate} } \geq 0 .
\end{align}
Equation (\ref{eq:microsecondlaw}) constitutes the second law of thermodynamics.
\subsection{Mesoscopic Thermodynamics}
The exact coarse-graining of the microscopic dynamics from above does not imply that the statistics of the thermodynamic observables are invariant under this marginalization \cite{esposito2012pre}. We define $\mesoenergy_{\mesostate}$ to be the internal energy of the system in the mesostate $\mesostate$ and find for the first law of thermodynamics 
\begin{align}
\d_t \langle \mesoenergy \rangle = \sum\limits_{\mesostate, \mesostate'} \mesoenergy(\mesostate) \, \mesorates_{\mesostate \mesostate'} \, \mesoprobability_{\mesostate'} = \langle \dot{\mesoheat} \rangle + \langle \dot{\mesowork} \rangle ,
\end{align}
with the mesoscopic heat and work currents
\begin{align} \label{eq:mesoheat}
\langle\dot{\mesoheat}\rangle &=  \sum\limits_{\mesostate, \mesostate'}  \big( \mesoenergy(\mesostate) - \force \, \mesosign  \big) \mesorates_{\mesostate \mesostate'}  \mesoprobability_{\mesostate'}   \\
\langle\dot{\mesowork}\rangle &= \sum\limits_{\mesostate, \mesostate'} f \, \mesosign  \mesorates_{\mesostate \mesostate'} \, \mesoprobability_{\mesostate'}  . \label{eq:mesowork}
\end{align}
Furthermore, we define the system entropy in the mesospace
\begin{align}  \label{eq:mesosystementropy}
\left\langle \mesoentropy \right \rangle &=  \sum\limits_{\mesostate} \mesoprobability_{\mesostate} \left( \multiplicity(\mesostate) - \ln \mesoprobability_{\mesostate} \right) \, ,
\end{align}
consisting of the non-equilibrium entropy defined by Eq. (\ref{eq:microsystementropy}) and the equilibrium entropy from Eq. \eqref{eq:internalentropy} accounting for the internal structure of the mesostates. Again, we split the time-derivative of the mesoscopic entropy into the mesocopic entropy flow 
\begin{align} \label{eq:mesoentropyflow}
\langle \dot{\mesoentropy}_e \rangle = -  \sum\limits_{\mesostate, \mesostate'} \mesorates_{\mesostate \mesostate'} \, \mesoprobability_{\mesostate'} \ln \frac{ {\microrates}_{\mesostate \mesostate'}}{ \microrates_{\mesostate' \mesostate}}
= \invtemperature \langle \dot{\mesoheat} \rangle ,
\end{align}
and the mesoscopic EP rate
\begin{align} \label{eq:mesosecondlaw}
\langle \dot{\mesoep} \rangle =  \sum\limits_{\mesostate, \mesostate'} \mesorates_{\mesostate \mesostate'} \, \mesoprobability_{\mesostate'} \, \ln \frac{\mesorates_{\mesostate \mesostate'} \mesoprobability_{\mesostate'} }{\mesorates_{\mesostate' \mesostate} \mesoprobability_{\mesostate} } \geq 0 .
\end{align}
that constitutes the second law of thermodynamics.
A closer inspection shows that while the first-law quantities are preserved under the coarse-graining procedure
\begin{align}
\d_t \langle \mesoenergy \rangle = \d_t \langle \microenergy \rangle, \quad \langle\dot{\mesoheat}\rangle = \langle\dot{\microheat}\rangle , \quad \langle\dot{\mesowork}\rangle = \langle\dot{\microwork}\rangle  ,
\end{align} 
the definitions in Eqs. \eqref{eq:mesosystementropy},\eqref{eq:mesosecondlaw} are in general not equivalent to the microscopic ones, \idest $\langle \mesoentropy \rangle \neq \langle \microentropy \rangle, \langle \mesoep \rangle \neq \langle \microep \rangle $. Yet, in the stationary state it holds that
$ \mesoprobability_{\mesostate} = \microprobability_{\microstate} \cdot \Omega(\mesostate) $, which in turn implies that the entropies in mesoscopic representation are identical to those in microscopic representation, \idest $\langle \mesoentropy^{s} \rangle = \langle \microentropy^{s} \rangle, \langle \mesoep^{s} \rangle = \langle \microep^{s} \rangle $. For this particular case, the second law
\begin{align} \label{eq:mesosecondlawstationary}
\!\!\!\! 
\langle \dot{ \mesoep^s } \rangle  \! = \! \sum\limits_{\mesostate, \mesostate'} \!\! \mesorates_{\mesostate \mesostate'} \, \mesosteadyprobability_{\mesostate'} \, \ln \frac{ \tilde{\mesorates}_{\mesostate \mesostate'}}{\tilde{\mesorates}_{\mesostate' \mesostate}} 
= - \langle \dot{\mesoentropy}_e^s  \rangle \geq 0  ,
\end{align}
states that the steady mesoscopic entropy flow is equal to minus the stationary mesoscopic entropy production rate.\\
As shown above, the mean-field Eq. \eqref{eq:meanfieldmasterequationderivation} results from a mean-field approximation applied on the mesoscopic ME. Thus, only the definitions of the thermodynamic observables in the mean-field limit stated in the main body are representing the true physical observables, if the corresponding microscopic definition coincides with the mesoscopic one. It therefore holds for the mean-field observables $\mathcal{X}$ that $ \lim_{\dimension \to \infty} \; \langle\dot{X}\rangle /\dimension = \dot{\mathcal{X}},$ with the mesoscopic observables $X \!\! ~=~ \!\! \mesoenergy,\mesoheat,\mesowork,\mesoentropyflow,\occupation_i $ and $ \lim_{\dimension \to \infty} \; \langle\dot{\mesoep}^{s}\rangle /\dimension = \dot{\mathcal{S}_i^{s}} $, where the superscript $^{s}$ refers to a stationary state.

\bibliography{bibliography}

\end{document}